# The Calculations of Gravity Fields and Rotation Curves of Whirlpool Galaxies and Dark Material

Mei Xiaochun ( Institute of Innovative Physics in Fuzhou, Fuzhou, China )
Xu Kuan ( Institute of Nuclear and New Energy Technology, Tsinghua University, Beijing )
Yu Ping (Cognitech Computer Inc., C A, U.S.A )

**Abstract** The gravity fields and rotation curves of whirlpool galaxies with thin disc distribution of material are calculated numerically. It is proved that the gravity field of mass thin disc distribution is greatly different from that of spherically symmetrical distribution. As long as the Newtonian theory of gravity is used strictly, by the proper mass distributions of thin discs, the flat rotation curves of whirlpool galaxies can be explained well. The rotating curve of the Milky Galaxy is obtained which coincides with practical observation. In this way, it is unnecessary for us to suppose the existence of additional dark material in the illuminant discs of whirlpool galaxy again. Meanwhile, in the space outside the illuminant disc, the quantity of dark material needed to maintain the flatness of rotation curves is greatly decreased. By considering the observation fact that the quantity of non-luminous baryon material is 3~10 times more than luminous material, we can explain the flatness of rotation curves of whirlpool galaxies well without the hypotheses of non-baryon dark material. So it is unnecessary for us to suppose that non-baryon dark material is about 5 times more than baryon material in a single whirlpool galaxy, no mater whether non-baryon dark material exits or not.

The gravity fields, rotating curves and dark material of whirlpool galaxies are discussed in this paper. It is proved by numerical calculation that the gravity field of mass thin disc distribution is greatly different from that of spherically symmetrical distribution. By using the Newtonian theory of gravitation strictly and considering the proper distribution of thin disc mass (including some baryon dark material), the flat rotation curves of whirlpool galaxies can be explained well without the existence of non-baryon material. That is to say, we can explain the flatness of rotation curves of whirlpool galaxies well based on the observation fact that the quantity of non-luminous baryon material is about 3~10 times of luminous baryon material. It is unnecessary for us to suppose that non-baryon dark material is about 5 times more than baryon material in whirlpool galaxies, no mater whether non-baryon dark material exits or not. The paper does not involve the problem of dark material existing among whirlpool galaxies and whirlpool galaxy groups. The conclusion about dark material is only for a single whirlpool galaxy.

We discuss the gravitational field of uniform infinite thin disc at first. Then discuss the gravity field and the rotation curve of the Milky Galaxy. At last discuss the problem of dark material in general whirlpool galaxies.

## 1. The gravity field of uniform and infinite thin disc

As well-known according to the Newtonian theory of gravity, a body moves steadily around a mass sphere in which material is distributed uniformly, the relation between the speed $V$ and the distance $r$

of body is

$$\frac{V^2}{r} = \frac{GM}{r^2} = F \qquad \text{or} \qquad V = \sqrt{rF} = \sqrt{\frac{GM}{r}} \qquad (1)$$

Suppose that the mass of a common galaxy with 100 billion sums is $M = 2 \times 10^{41} kg$, which is distributed uniformly in a sphere with a radius of $5.37 kpc$ (about $16.6 \times 10^{20} m$). In light of (1), we can obtain the rotation rotating curve $L_1$ as shown in Fig 3. Insider the sphere, the rotation curve goes up slowly. Outsider the sphere, the curve goes down quickly. On the other side, as we known that the shape of whirlpool galaxies is compressed like a disc. However, it is difficult to calculate the gravitational field of material disc distribution. We can not obtain the analytical form of function. In the current celestial mechanics, we generally calculate the gravity potential of mass thin disc, in which the ellipse function is involved. Then we obtain the gravity fields by differential calculation. Because of mathematical difficulty, only a few disc gravity potentials are obtained. For example, when material density decays in the form of exponent, we can obtain the gravity potential of disc [1]. Based on it, the rotation curve can be drawn, of which the shape is similar to that based on (1). So at present, we always use (1) to estimate the rotation curves of common whirlpool galaxies without considering the influence of disc shapes of whirlpool galaxies on the rotation curves.

However, the practical observations found that the rotation curves were nearly flat with the relation $V \approx$ constant, i.e., the rotating speeds of celestial bodies in whirlpool galaxies had nothing to do the distances between the celestial bodies and the centers of whirlpool galaxies. Fig. 1 shows the some rotating curves of whirlpool galaxies [2]. In order to explain the flat characteristics of rotation curves, the hypotheses of dark material was proposed. There are two kinds of dark material. One is baryon dark material including interstellar thin gas, planets and dead starts and so on. Another is non-baryon dark material of which nature are unknown expect they have no electromagnetic interaction. Estimated by the theory of cosmology, non-baryon dark material is about 5 times of baryon dark material in the universe. It is a basic fact that there exists dark material in the universe. In light of the astronomical observations, the quantity of non-luminous baryon material is about $3 \sim 10$ times more than luminous baryon material in galaxies [3]. But non-baryon dark material is still a hypothesis at present. Though astronomers have strived for more than 30 years, they can not make certain that where or not non-baryon dark material exist. The problem of non-baryon dark material has become one of biggest enigmas in modern astrophysics and cosmology.

On the other hand, astronomical observation shows that the moving speeds of celestial bodies in spherical galaxies still satisfy (1) without the relation $V \approx$ constant. So it is unnecessary for us to suppose the existence of large amount of non-baryon dark material in the spherical galaxies. In this way, a problem is raised, i.e., why so much non-baryon dark material exists in whirlpool galaxies but not exist in spherical galaxies. In the fact, both kinds of galaxies have no essential difference. This fact actually hints us whether or not it is improper to use (1) to estimate the rotation speeds of whirlpool galaxies. It is proved below that the gravitation of mass thin disc distribution is quite different from that of spherical distribution. As long as we calculate the gravity fields of whirlpool galaxies based on the Newtonian theory strictly, we can obtain the nearly flat rotation curves, so that it is unnecessary for us to suppose the existence of additional dark material in the luminous discs of whirlpool galaxies. The quantity of dark material needed to maintain the flatness of rotation curves outsider the luminous discs is also decreased greatly. According to the practical observation for the ratio of mass and luminosity, non-luminous baryon dark material is about $3 \sim 10$ times of luminous baryon dark material in general galaxies. Based on this ratio, we can explain the flatness of rotation curves outsider the luminous discs of whirlpool galaxies. Therefore, it is unnecessary for us to suppose that non-baryon dark material is about 5 times more than baryon material in a single whirlpool galaxy, no mater whether non-baryon dark material exits or not.

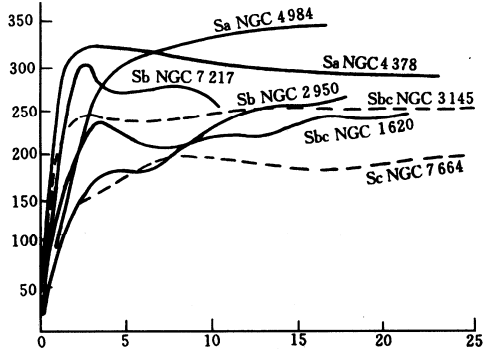 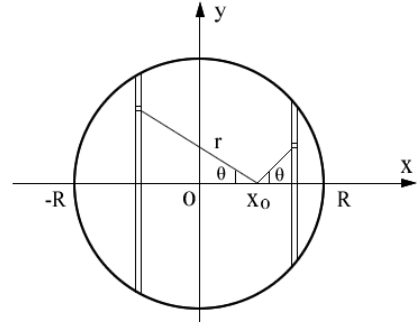

**Fig. 1 Rotating curves of some whirlpool galaxies**     **Fig. 2 Gravity field of uniform thin disc**

For simplification at first, we discuss the gravity field of mass infinite thin disc distribution. In stead of calculating gravity potential, we calculate gravity directly. As shown in Fig. 2, suppose that the radius of disc is $R$, the thickness is neglected and the surface density of mass is a constant $\sigma_0$. The gravity between the mass of area element $dxdy$ and the unite mass located at point $x_0$ is

$$d\bar{F}(x_0) = -\frac{G\sigma_0 dxdy\, \bar{r}}{r^3} \qquad r = \sqrt{(x_0-x)^2 + y^2} \qquad (2)$$

By the consideration of symmetry, the partial force in the direction of $y$ axis is zero. The differential gravity along the direction of $x$ axis is

$$dF(x_0) = -\frac{G\sigma_0 \cos\theta\, dxdy}{r^2} \qquad \cos\theta = \frac{x_0 - x}{\sqrt{(x_0-x)^2 + y^2}} \qquad (3)$$

The gravity of while disc acted on the unit mass located at point $x_0$ is

$$F(x_0) = -\int_{-R}^{R} dx \int_{-\sqrt{R^2-x^2}}^{\sqrt{R^2-x^2}} \frac{G\sigma_0 (x_0 - x)dy}{\left((x_0-x)^2 + y^2\right)^{3/2}} = -\int_{-R}^{R} \frac{2G\sigma_0 \sqrt{R^2 - x^2}\, dx}{(x_0 - x)\sqrt{R^2 + x_0^2 - 2x_0 x}} \qquad (4)$$

When $x_0 > R$, we have $x_0 - x > 0$. It is easy to prove that we have $R^2 + x_0^2 - 2x_0 x > 0$ in this case. So in the region $x_0 > R$, the integrated function is normal. But in the region $x_0 \leq R$, the integrated function is infinite at point $x = x_0$, This mathematical difficulty may be the reason why we does not understand the nature of thin disc gravity field well up to now.

We now prove that the integral infinite doe not exist actually in the disc region with $x_0 < R$. The infinite only appears at the edge point $x_0 = R$. Then we prove that the infinite at the edge point can be eliminated by choosing proper mass distribution so that the mass density becomes zero at the edge of thin disc. The formula (4) is actually non-integrable. But we can divide it into three parties. Let $\Delta$ be a small quantity, we have

$$F(x_0) = -\int_{-R}^{x_0-\Delta} \frac{2G\sigma_0\sqrt{R^2-x^2}\,dx}{(x_0-x)\sqrt{R^2+x_0^2-2x_0x}} - \int_{x_0+\Delta}^{R} \frac{2G\sigma_0\sqrt{R^2-x^2}\,dx}{(x_0-x)\sqrt{R^2+x_0^2-2x_0x}}$$

$$-\int_{x_0-\Delta}^{x_0+\Delta} \frac{2G\sigma_0\sqrt{R^2-x^2}\,dx}{(x_0-x)\sqrt{R^2+x_0^2-2x_0x}} \tag{5}$$

There are no singularities in the first and third items on the right side of (5). The singularity exists in the second item, so we only discuss the second item and have

$$F_3(x_0) = \int_{x_0-\Delta}^{x_0+\Delta} \frac{2G\sigma_0\sqrt{R^2-x^2}\,dx}{(x_0-x)\sqrt{R^2+x_0^2-2x_0x}}$$

$$= \int_{x_0}^{x_0+\Delta} \frac{2G\sigma_0\sqrt{R^2-x^2}\,dx}{(x_0-x)\sqrt{R^2+x_0^2-2x_0x}} - \int_{x_0}^{x_0-\Delta} \frac{2G\sigma_0\sqrt{R^2-x^2}\,dx}{(x_0-x)\sqrt{R^2+x_0^2-2x_0x}} \tag{6}$$

Let $x = x_0 + u$ in the first item and $x = x_0 - u$ in the second item in (6), we write the formula as

$$F_3(x_0) = \int_0^\Delta \frac{2G\sigma_0\sqrt{R^2-x_0^2-2x_0u-u^2}\,du}{-u\sqrt{R^2-x_0^2-2x_0u}} - \int_0^\Delta \frac{2G\sigma_0\sqrt{R^2-x_0^2+2x_0u-u^2}\,d(-u)}{u\sqrt{R^2-x_0^2+2x_0u}}$$

$$= -2G\sigma_0 \int_0^\Delta \frac{du}{u}\sqrt{1-\frac{u^2}{R^2-x_0^2-2x_0u}} + 2G\sigma_0 \int_0^\Delta \frac{du}{u}\sqrt{1-\frac{u^2}{R^2-x_0^2+2x_0u}} \tag{7}$$

Because $\Delta$ is a small quantity, $u$ is also a small quantity in the integral region, we have the Taylor's series development

$$\sqrt{1-\frac{u^2}{R^2-x_0^2-2x_0u}} = 1 - \frac{u^2}{2(R^2-x_0^2-2x_0u)} + \cdots \tag{8}$$

$$\sqrt{1-\frac{u^2}{R^2-x_0^2+2x_0u}} = 1 - \frac{u^2}{2(R^2-x_0^2+2x_0u)} + \cdots \tag{9}$$

The higher order items in (8) and (9) can be neglected so that (7) can be written as approximately

$$F_3(x_0) = 2G\sigma_0 \int_0^\Delta \frac{du}{u} \left( \frac{u^2}{2(R^2-x_0^2+2x_0u)} - \frac{u^2}{2(R^2-x_0^2-2x_0u)} \right)$$

$$= -2G\sigma_0 \int_0^\Delta \frac{2x_0 u^2\,du}{(R^2-x_0^2)^2 - (2x_0u)^2}$$

$$= \frac{G\sigma_0\Delta}{x_0} - \frac{G\sigma_0(R^2-x_0^2)^2}{x_0} \int_0^\Delta \frac{du}{(R^2-x_0^2)^2-(2x_0u)^2}$$

$$= \frac{G\sigma_0\Delta}{x_0} + \frac{G\sigma_0(R^2-x_0^2)}{4x_0^2} \left[ \ln\frac{R^2-x_0^2-2x_0\Delta}{R^2-x_0^2+2x_0\Delta} \right] \tag{10}$$

When $x_0 \neq R$ and $\Delta \to 0$, we have $F_3(x_0) \to 0$. So (5) is finite when $x_0 \neq R$.

On the other hand, (4) is integrable at point $x_0 = R$. Let $R - x_0 = z$ in (4), we consider integral

$$F(R) = -\frac{2G\sigma_0}{\sqrt{2R}} \int_{-R}^{R-\Delta} \frac{\sqrt{R+x}\,dx}{(R-x)} = -\frac{2G\sigma_0}{\sqrt{2R}} \int_{\Delta}^{2R} \frac{\sqrt{2R+z}\,dz}{z}$$

$$= -\frac{2G\sigma_0}{\sqrt{2R}} \left[ 2\sqrt{2R+z} \Big|_{\Delta}^{2R} + 2R \int_{\Delta}^{2R} \frac{dz}{z\sqrt{2R+z}} \right] \quad (11)$$

By means of the formula

$$\int_{\Delta}^{2R} \frac{dz}{z\sqrt{2R-z}} = \frac{1}{\sqrt{2R}} \ln \frac{\sqrt{2R}-\sqrt{2R-z}}{\sqrt{2R}+\sqrt{2R-z}} \Big|_{\Delta}^{2R} = \frac{1}{\sqrt{2R}} \left[ \ln \frac{\sqrt{2R}}{\sqrt{2R}} - \ln \frac{\sqrt{2R}-\sqrt{2R-\Delta}}{\sqrt{2R}+\sqrt{2R-\Delta}} \right]$$

$$\approx \frac{1}{\sqrt{2R}} \left[ \ln 1 - \ln \frac{\sqrt{2R}-\sqrt{2R}(1-\Delta/4R)}{\sqrt{2R}+\sqrt{2R}(1-\Delta/4R)} \right] \approx -\frac{1}{\sqrt{2R}} \ln \frac{\Delta}{8R} \quad (12)$$

We get

$$F(R) = -\frac{2G\sigma_0}{\sqrt{2R}} \int_{-R}^{R-\Delta} \frac{\sqrt{R+x}\,dx}{(R-x)} = -\frac{2G\sigma_0}{\sqrt{2R}} \int_{0}^{2R-\Delta} \frac{\sqrt{2R+z}\,dz}{z}$$

$$\approx -2G\sigma_0 \left[ 2(\sqrt{2}-1) - \ln \frac{\Delta}{8R} \right] \quad (13)$$

When $\Delta \to 0$, (13) is just (4), but the integral is infinite. However, practical gravity can not be infinite at the edge of thin disc. We prove below that the infinite can be eliminated if we suppose that the mass density decreases with the increase of radius and becomes zero at the edge of thin disc.

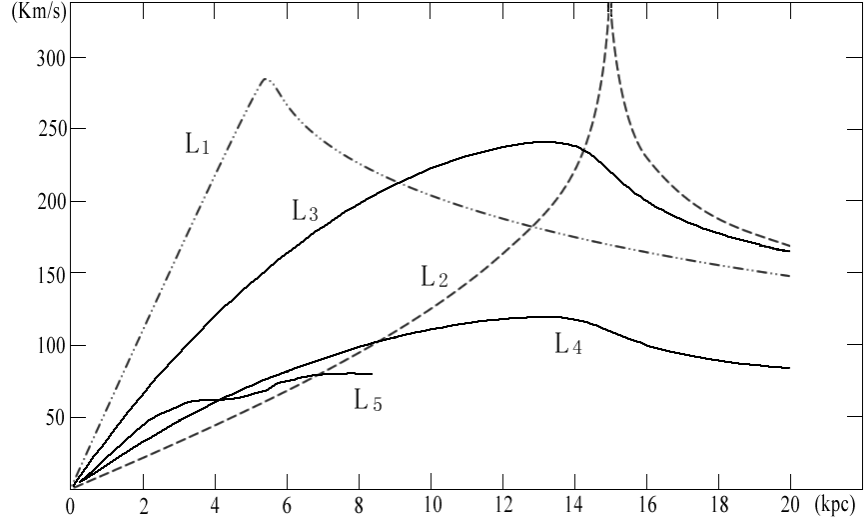

**Fig. 3.** $L_1$ is the rotation curve of uniform mass sphere. $L_2$ is the rotation curve of uniform and infinite thin disc. $L_3$, $L_4$ are the rotation curves of thin disc of which densities linearly decrease. $L_5$ is the rotation curve of NGC1560 whirlpool galaxy

Suppose that a whirlpool galaxy contains 100 billion suns with mass $M_1 = 2 \times 10^{41} Kg$ (including luminous and non-luminous baryon material). The radius of galaxy disc is 49 thousand light-years, i.e., $R_1 = 4.46 \times 10^{20} m = 15 kpc$. At first, we consider the thickness of galaxy disc is 2 thousand light-years, i.e., $h_1 = 1.89 \times 10^{19} m = 0.61 kpc$. The volume density of disc is $\rho_1 = 7.85 \times 10^{-21} Kg/m^3$. Then considering the whirlpool galaxy as an infinite thin disc again, we can get that the surface density of disc is $\sigma = M_1/(\pi R_1^2) = 0.30 Kg/m^2$. By the numerical calculation of (4) using computer and considering the relation $V = \sqrt{x_0 F}$, we obtain the rotation curve $L_2$ as shown in Fig. 3. In light of $L_2$, the gravity and rotation speed increase slowly with the increase of $x_0$. At the edge of thin disc with $x_0 = 15 kpc$, the rotation speed reaches its biggest value (infinite). At the point $x_0 = 20 kpc$ outside the disc, the rotation speed is $V = 170 km/s$. On the other hand, if considering the galaxy as a uniform sphere with the same mass density $\rho = 7.85 \times 10^{-21} Kg/m^3$, the radius of sphere is $5.37 kpc$. In light of (1), we obtain the rotation curve $L_1$ shown in Fig.3. At the edge of sphere with $x_0 = 5.37 kpc$, the rotation speed reaches its biggest with $V = 284 km/s$. Then the rotation speed decreases quickly. In the position $x_0 = 15 kpc$, we have $V = 170 km/s$. In the position $x_0 = 20 kpc$, we have $V = 146 km/s$. It is obvious that for the same mass and density, curves $L_2$ and $L_1$ are very different. The peak of curve $L_2$ is located at the right side of the peak of curve $L_1$ much far. In the region outside the disc, the rotation speed of $L_2$ is higher then that of $L_1$. These two factors would greatly influence the rotation curve's shapes of practical whirlpool galaxy and the quantity of dark material needed to maintain the flatness of rotating curves.

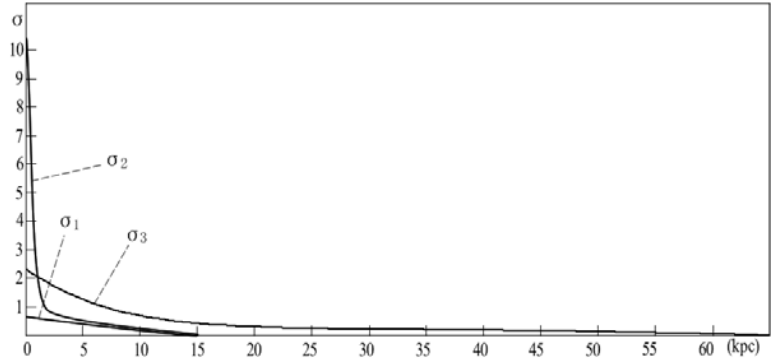

**Fig. 4 The curves of mass surface densities of whirlpool galaxies**

In the general whirlpool galaxies, mass surface density is not a constant. Suppose that the mass surface density decreases linearly in the form below

$$\sigma(r) = \sigma_0(1 - 0.067r) = \sigma_0(1 - 0.067\sqrt{x^2 + y^2}) \tag{14}$$

Here $r = \sqrt{x^2 + y^2}$ is the distance between a certain point and the center of disc. By taking $kpc$ as the unite of length, the density becomes zero at the edge of disc with $r = 15 kpc$. Suppose that the mass of a whirlpool galaxy is still $2 \times 10^{41} Kg$, in light of (14), we can calculate and get $\sigma_0 = 0.89 Kg/m^2$. The function relation of (14) is described by curve $\sigma_1$ in Fig. 4. Substitute (14) into (4), we get

$$F(x_0) = -\int_{-R}^{R} dx \int_{-\sqrt{R^2-x^2}}^{\sqrt{R^2-x^2}} \frac{G\sigma_0(1 - 0.067\sqrt{x^2+y^2})(x_0-x)dy}{\left((x_0-x)^2 + y^2\right)^{3/2}} \tag{15}$$

The direct integral is also difficult. By the numerical calculation of computer, we obtain curve $L_3$ shown in Fig.3. The curve raises slowly and reaches its highest point at the point $14 kpc$, then bended down slowly. The infinite disappears. If the mass of disc is $M_2 = M_1/4 = 5 \times 10^{40} kg$, we obtain curve $L_4$ of which front part is similar to the rotation curve $L_5$ of NGC1560 whirlpool galaxy. The character of both curves is that rotation speed increases monotonously with the increase of distance. As long as enhancing mass density in the center region of disc, we can use improved $L_4$ to describe the rotation curve of NGC1560 whirlpool galaxy approximately.

## 2. The rotation curve and dark material of the Milky Galaxy

There is a ridged core in the center of common whirlpool galaxy in general. The density of core is much high than that on the disc. As we known that the core radius of the Milky Galaxy is about $2 \sim 3kpc$ and the radius of disc is about $12.5 \sim 15kpc$. Suppose that $0.72 \times 10^{40} kg$ material is distributed uniformly in the sphere core of galactic disc with radius $0.25kpc$, $1.48 \times 10^{40} kg$ material is distributed uniformly in the spherical shell with radius $1 \sim 2kpc$, $10^{40} kg$ is distributed uniformly in the spherical shell with radius $1 \sim 2kpc$. Then $0.6 \times 10^{40} kg$ is distributed uniformly in the spherical shells with radius $2 \sim 3kpc$, $3 \sim 4kpc$ and $4 \sim 5kpc$ individually. That is to say, total mass $M_1 = 5 \times 10^{40} kg$ is distributed in the sphere with radius $5kpc$. By transforming the volume densities into surface densities, the relation between surface density and distance is described by $\sigma_2$ shown in Fig. 4. The gravity acted on the unite mass located at the point $x_0$ can be written as

$$F(x_0) = F_s(x_0) - \int_{-R}^{R} dx \int_{-\sqrt{R^2-x^2}}^{\sqrt{R^2-x^2}} \frac{G\sigma_0(1 - 0.067\sqrt{x^2+y^2})(x_0-x)dy}{((x_0-x)^2 + y^2)^{3/2}} \tag{16}$$

In which $F_s(x_0)$ is the additional gravity caused by the mass sphere with radius $5kpc$. Based on (16) and by the numerical calculation, we obtain rotation curve $L_7$ shown in Fig. 5, which is similar to the practical rotation curve $L_6$ of the Milky Galaxy. The character of $L_6$ that in the center of galaxy core, the rotating speed increase quickly, then decreases, then increase again. This is different from that of NGC1560 whirlpool galaxy. In the front parts, $L_6$ and $L_7$ almost overlap. In the region of $8 \sim 15kpc$ $L_7$ is little higher than $L_6$ and in the region $15 \sim 20kpc$ outside the disc, $L_7$ is lower than $L_6$. At the edge point $15kpc$, two curves meet with same rotation speeds. So inside the luminous disc of the Milky Galaxy with radius $r \leq 15kpc$, we need no additional dark material to explain the nearly flat rotation curve. The curve $L_8$ represent the rotation curve when $2 \times 10^{41} kg$ mass is distributed uniformly in a sphere with radius $2.5kpc$ according to (1). It is obvious that the difference are great between $L_6$, $L_7$ and $L_8$. It indicates that there exists great error for us to use the spherical distribution of mass to replace the disc distribution to estimate the rotation curves of whirlpool galaxies. Therefore, it is completely improper. In stead of using spherical mode, as long as we calculate the disc gravitational fields of whirlpool galaxies directly, the additional dark material would become unnecessary to maintain the flatness of rotation curves inside the luminous discs. Then let's estimate how much dark material is need to raise the tail of $L_7$, so that it can be coincide with the practical observation curve $L_6$.

As we known that there exists much faint material in the space outsider the luminous discs of whirlpool galaxies. The faint material includes a few stars and dark material just as interstellar gas and dead starts and so on. According to cosmology, faint material would include non-baryon dark material too. If the average material density of the current universe is taken as $\Omega = 1$, in light of current cosmology, the density of dark energy is $\Omega_e \approx 0.7$, the average density of non-baryon dark material is $\Omega_n \approx 0.25$ and the average density of baryon material is only $\Omega_b \approx 0.05$. In light of practical observation for the ratio of mass and luminosity in astronomy, the quantity of non-luminous baryon material is about 3 ~ 10 times of luminous baryon material. Because the distribution region of faint material outside the luminous area is very large, it is acceptable to think that the quantity of non-luminous baryon material is about 9 ~ 10 times of luminous baryon material in general galaxies. So the average density of luminous baryon material may be about $\Omega_l \approx 0.005$ and the average density of non-luminous baryon material may be about $\Omega_d \approx 0.045$. In this estimation, the quantity of non-baryon dark material may be about 50 times of luminous baryon material.

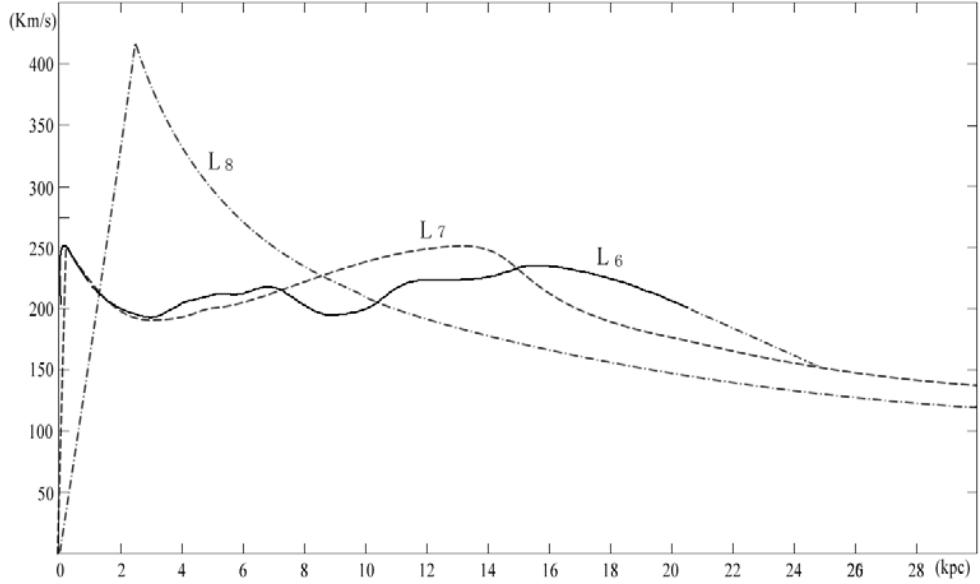

**Fig. 5** $L_6$ is the observed rotation curve of the Milky Galaxy. $L_7$ is the theoretical rotation curve of the Milky Galaxy. $L_8$ is the rotation curve when the mass of the Milky Galaxy is concentrated in the spherical core with radius $2.5kpc$.

Based on this ratio, we can estimate the quantities of baryon and non-baryon dark material in the Milky Galaxy through fig. 5. In light of practical observation curve $L_6$, we know that the rotation speed is $V_r = 206 Km/s$ at the point $20kpc$ outside the luminous disc. According to the theoretical curve $L_7$, the rotation speed is $V_c = 178 Km/s$ at the same point. If there exists faint dark material outside the disc, the rotation speed of $L_7$ can be raised. The distribution of faint dark material outsider the disc is considered with spherical symmetry. Suppose that the mass $M_d$ of faint material is uniformly distributed between two sphere shells with thickness $x_{20} - x_{15} = 5kpc$, so that the theoretical rotation speed at the point $x_0 = 20kpc$ can be raised to $V_r = 206 Km/s$. In light of the Newtonian theory of gravity, this kind of mass distribution does not affect the rotating speed of body inside the spherical shell, for the gravity caused by the mass spherical shell is zero inside the shell. When the additional gravity acted on the body at the point outside the shell is calculated, the mass can considered to be concentrated at the center of the spherical shell. In this way, we have relation

$$\frac{V_r^2}{x_{20}} = \frac{V_c^2}{x_{20}} + \frac{GM_d}{x_{20}^2} = \frac{V_c^2}{x_{20}} + \frac{4\pi G \rho_d (x_{20}^3 - x_{15}^3)}{3x_{20}^2} \tag{18}$$

From the formula, we can obtain $M_d = 9.96 \times 10^{40} Kg$, which is about $0.5$ time of the luminous disc's mass. The density of faint dark material is $\rho_d = 1.74 \times 10^{-22} Kg/m^3$. As said before, the volume density of luminous disc $\rho = 7.85 \times 10^{-21} Kg/m^3$, which is about 45 times of $\rho_d$. So the density of faint dark material is quite thin. In the document (2), the rotation curve of the Milky Galaxy ends at the point $20.4 kpc$. In light of the trend, the observed curve $L_6$ would meet the theoretical curve $L_7$ at the point $25 kpc$. So for the Milky Galaxy, we do not need additional dark material in the luminous disc. In the space outside the disc, we do not need too much dark material. This kind of dark material can be considered as baryon material. That is to say, we need no the hypotheses of non-baryon dark material to explain the rotation curve of the Milky Galaxy actually. Some amount of baryon dark material is enough.

## 3. The rotation curves and dark material of general whirlpool galaxies

When the faints material is considered, the practical extents of whirlpool galaxies may be far large than the luminous disc. The extents of general whirlpool galaxies is considered with the magnitude $1 \sim 100 kpc$. New let's discuss a bigger whirlpool galaxy. For simplification, we use the disc distribution to substitute the spherical distribution for faint baryon dark material outside the disc. Take the radius of the galaxy $R = 64.7 kpc = 2.01 \times 10^{21} m$ and suppose that the region with radius $r \leq 20 kpc$ is the luminous disc. The other part is the region containing baryon dark material. Let $\alpha = 1/R = 0.015 (kpc)^{-1}$, $r = \sqrt{x^2 + y^2}$, the mass surface density to be (including luminous material and dark material)

$$\sigma(r) = 0.30(1 - \alpha r) + 2e^{-10\alpha r} = 0.30(1 - 0.015r) + 2e^{-0.15r} \tag{19}$$

In the formula, the first item describes linear decay of density, and the second describe the exponential decay which can quick concentrate the mass to the center of galaxy. The relation of density and distance is described by curve $\sigma_3$ in Fig. 4. By the numerical calculation, we get the rotation curve $L_9$ in Fig. 6. It indicates that when exponential decay is considered, the rotation curve becomes more flat than $L_3$ in Fig. 3. If much more mass is moved to the center of galaxy, similar to the rotation curve $L_7$ of the Milky Galaxy, $L_9$ can be reconstructed to describer the practical rotation speed of JGC2885 whirlpool galaxy[1] described by $L_{13}$ in Fig. 6. The mass of galaxy described by the rotating curve $L_9$ is estimated below

$$M = \int_{r_1}^{r_2} 2\pi \sigma(r) r dr = 2\pi \int_{r_1}^{r_2} \left(0.30(1 - 0.015r) + 2e^{-0.15r}\right) r \, dr$$

$$= 0.30\pi \left(r^2 - 0.01r^3\right) \Big|_{r_1}^{r_2} - 26.67\pi \, e^{-0.225r} \left(r + 6.67\right) \Big|_{r_1}^{r_2} \tag{20}$$

In which $kpc$ is taken as the unit of length. Let $r_1 = 0$, $r_2 = 65 kpc$, we have $M = 18.63 \times 10^{41} Kg$. By the same method, the mass of luminous disc with radius $20 kpc$ is $M_1 = 7.15 \times 10^{41} Kg$ and the mass of faint baryon material outside the luminous disc is $M_2 = 11.48 \times 10^{41} Kg$. Therefore, $M_2$ is about 1.6 times of $M_1$. In stead of the disc distribution, because the faint baryon material is distributed spherically, much more mass is needed to maintain the flatness of rotation curve, so the baryon dark material of $11.48 \times 10^{41} Kg$ can be considered as the lower limit to maintain the curve $L_9$. According to current cosmology, the mass of non-baryon dark material would be 5 times more than that of baryon material in general galaxies. Therefore, there would exist simultaneously more than $93.15 \times 10^{41} Kg$ non-baryon dark material between two spherical shells with radius $20 \sim 65 kpc$ outside the luminous disc. The density of non-baryon dark material would be $2.82 \times 10^{-22} Kg/m^3$. If there exists so much dark material additionally in the galaxy, rotation curve $L_9$ would become $L_{10}$ in Fig.6, which is not a nearly flat curve again. At the point $65kpc$, the rotating speed would be $V = 552 Km/s$, about 1.9 time of $L_9$. This is only the lower limit of rotation speed. By considering the fact that much more baryon material, therefore much more non-baryon dark material, is needed when their distribution is spherically symmetrical, the rotation curve would bend up much more. This result would not coincide with practical observation completely.

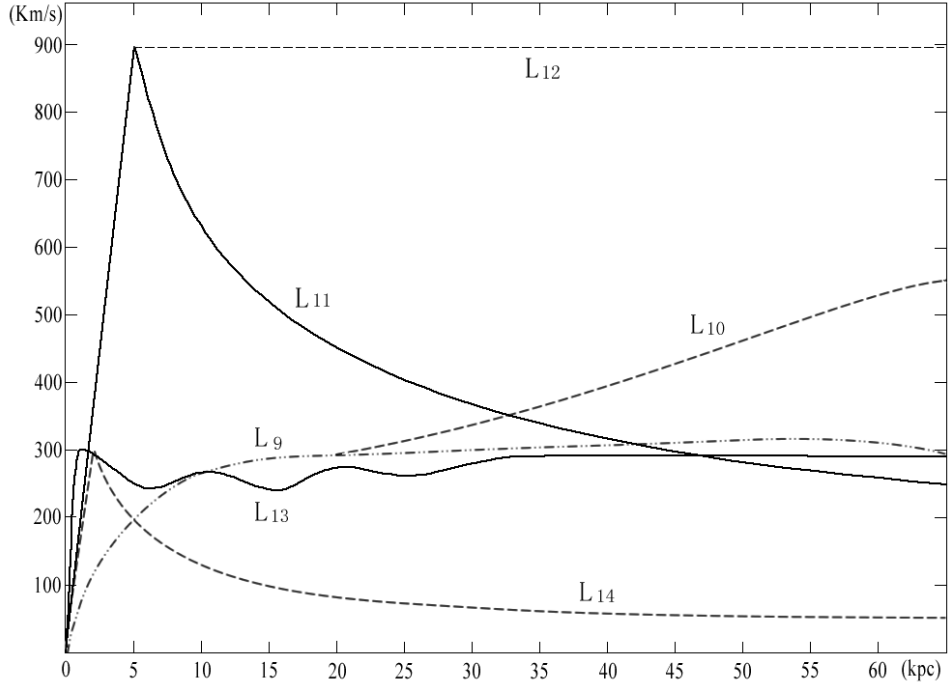

**Fig. 6** $L_9$ is the theoretical rotation curve of whirlpool galaxy containing baryon dark material without non-baryon dark material. $L_{10}$ is the theoretical rotation curve while non-baryon dark material is 5 times of baryon dark material. $L_{11}$ is the rotation curve when the mass of $18.63 \times 10^{41} Kg$ is concentrated in a sphere with radius $5kpc$. $L_{12}$ is the parallel line of the highest value of $L_{11}$. $L_{13}$ is the practically observed rotation curve of JGC2885 whirlpool galaxy. $L_{14}$ is the rotation curve while the mass $8.34 \times 10^{40} Kg$ is concentrated in a sphere with radius $2kpc$.

The curve $L_{11}$ in Fig. 6 is the rotating curve when mass of $18.63 \times 10^{41} Kg$ is concentrated in a sphere with radius $5kpc$. $L_{12}$ is the parallel line of the highest value of $L_{11}$ with the rotation speed $V_5 = 897 Km/s$. At the point $x_{65} = 65kpc$, the rotation speed of $L_{11}$ is $V_{65} = 248 Km/s$. In order to let the rotation curve of spherical distribution becoming flat, dark material is needed. We estimate the needed quantity of dark material in light of the formula below

$$\frac{V_5^2}{x_{65}} = \frac{V_{65}^2}{x_{65}} + \frac{GM_d}{x_{65}^2} \tag{21}$$

The result is that the dark material of $M_d = 2.24 \times 10^{43} Kg$ is needed in two spherical shells with radius

$5 \sim 65 kpc$. The mass is about 31 times of the luminous disc's mass and about 12 times of the total luminous and non-luminous faint material of the whirlpool galaxy. This amount of material can only be non-baryon dark material. This is just the reason why we need the hypotheses that there exists a large amount of non-baryon dark material in the whirlpool galaxies. But if we calculate the gravitational field of thin disc directly based on the Newtonian theory of gravity, what we obtain is the curve $L_9$. In light of $L_9$, we do not need so much non-baryon dark material.

What discussed above is the situations that when the mass of galaxy is known. For example, the Milky Galaxy, of which mass can be estimated by taking count of star's number. However, in most situations, the mass of galaxies can not be known certainly. We can only estimate their masses by their rotation curves. For example, the JGC2885 whirlpool galaxy, of which rotation curve is $L_{13}$ shown in Fig. 6. Far to $100 kpc$, the curve is still flat, but we do not know its mass. The curve $L_{13}$ has a distinct peak in its front part of $1.5 \sim 2 kpc$ with the rotation speed $V = 300 Km/s$. If (1) is used to estimate its mass, a great amount of non-baryon dark material would be needed to maintain the flatness of curve. In fact, suppose that the luminous and non-luminous material of a galaxy is concentrated in a sphere with radius $2 kpc$, the rotation speed is $V = 300 Km/s$ at the edge of the sphere. In light of (1), the mass of galaxy would be $M = 8.34 \times 10^{40} Kg$ and the rotation speed would become $L_{14}$ in Fig 6, which indicates that the rotation speed would be $V = 52.6 Km/s$ at the point of $65 kpc$. However, according to practical observation curve $L_{13}$, the rotating speed is $V = 290 Km/s$ at the point of $65 kpc$. So dark material is needed to keep the flatness of rotation curve in the sphere shells with radius $2 \sim 65 kpc$. Similar to the calculation of (16), the mass of needed dark material would be $2.45 \times 10^{42} Kg$, which is 29.4 times of the galactic mass $8.34 \times 10^{40} Kg$. This amount of dark material can only be non-baryon dark material, in stead of baryon dark material. Therefore, it is obvious that using spherical distributions to replace the disc distributions to estimate the rotation speeds of whirlpool galaxies would lead to high estimation of non-baryon material in whirlpool galaxies. This may be the reason that we suppose the existence of a large amount of non-baryon dark material in whirlpool galaxies, but can not find them up to now. If our Milky Galaxy has so much amount of non-baryon dark material really, it would be easy to find it in generally speaking.

## 4. Discussion

According to the Newtonian theory of gravity, when material is distributed uniformly in a sphere with radius $R$, the gravity caused by the sphere at the point $r < R$ is only relative to the mass contained inside the sphere shell with radius $r$. The result is equal to the situation that the mass is concentrated at the sphere center. The mass contained outsider the spherical shell does not produce gravity at the point insider the spherical shell. So the gravitational field caused by the spherically symmetric distribution of material is especially simple. However, the gravitational field caused by the disc distribution of material has no such simple nature. So the gravity caused by thin disc distribution is completely different from that of spherical distribution. Using the spherical distribution to replace the thin disc distribution to estimate the rotation curves of whirlpool galaxies would cause great error.

So when we discuss the rotation curves of whirlpool galaxies, we should calculate the gravity fields of thin discs directly in light of the Newtonian theory of gravity. The result is that inside the luminous disc of whirlpool galaxies, we need no additional dark material to explain the flat rotation curves. By considering the observation fact that the quantity of non-luminous baryon material is 3~10 times more than luminous material, we can also explain the flatness of rotation curves of whirlpool galaxies outside the luminous discs well. We should re-estimate the ratio of luminous baryon material, baryon dark material and non-baryon dark material in whirlpool galaxies. The hypotheses that a lot of non-baryon dark material exists in common whirlpool galaxies is unnecessary. Of cause, we can think that the dark material in whirlpool galaxies contains a little amount of non-baryon dark material. But it is unnecessary for us to suppose that non-baryon dark material is about 5 times more than baryon material in whirlpool galaxies. Otherwise, as shown by $L_{10}$ in Fig. 6, the rotation curves of whirlpool galaxies would bend up greatly. Such great amount of non-baryon dark material may exist in other places, but it can not exist in a single whirlpool galaxy just as our Milky Galaxy, though a little my do.

The authors are grateful to Dr. Lu Qinghua and An Longzhi, Institute of Innovative Physics in Fuzhou, for their careful calculation and great support on this paper.